\documentclass{article}
\usepackage{arxiv}

\usepackage[T1]{fontenc}
\usepackage{microtype}

\usepackage{graphicx}
\usepackage{xcolor}

\let\oldparagraph\paragraph
\renewcommand{\paragraph}[1]{\oldparagraph{{#1.}}}

\usepackage{amsmath,amssymb,amsthm,enumitem,mathtools}

\usepackage{booktabs}
\usepackage{adjustbox}
\usepackage{placeins}

\definecolor{linkblue}{HTML}{1F4E9A}
\usepackage[colorlinks=true,linkcolor=linkblue,citecolor=linkblue,urlcolor=linkblue]{hyperref}
\usepackage{orcidlink}
\hypersetup{pdftitle={Cryptanalysis of operad-based cryptography}, pdfauthor={Ambrogio Maria Bernardelli}}

\newcommand{\reals}{{\mathbf R}}
\newcommand{\integers}{{\mathbf  Z}}

\newtheorem{theorem}{Theorem}[section]  
\newtheorem{lemma}[theorem]{Lemma}  
\newtheorem{corollary}[theorem]{Corollary}
\newtheorem{proposition}[theorem]{Proposition}

\theoremstyle{definition}
\newtheorem{definition}[theorem]{Definition}

\theoremstyle{remark}
\newtheorem{example}[theorem]{Example}

\newcommand{\Zt}{\widetilde Z}
\newcommand{\pw}[2]{#1[#2]}          
\newcommand{\INT}{\mathsf{Int}}
\newcommand{\DIV}{\mathsf{Div}}
\newcommand{\Q}{\mathbf{Q}}

\title{Cryptanalysis of operad-based cryptography}
\author{Ambrogio Maria Bernardelli\,\orcidlink{0000-0002-2328-7062}\thanks{%
  Department of Mathematics ``F. Casorati'', University of Pavia,
  27100 Pavia, Italy.\newline
  \href{mailto:ambrogiomaria.bernardelli@unipv.it}{\texttt{ambrogiomaria.bernardelli@unipv.it}}}}
\date{}
\renewcommand{\shorttitle}{Cryptanalysis of operad-based cryptography}

\begin{document}
	\maketitle

	\begin{abstract}
		Commutative operads have been proposed as a platform for public-key cryptography, with protocols for key agreement, key exchange, encryption and authentication, together with a masking technique and families of algebras with secret parameters intended to strengthen them.
		We show that breaking key agreement, key exchange and encryption reduces to two problems over the \emph{centralizer} of the public operations: an interpolation problem for key agreement and a division problem for the other two.
		Neither problem requires recovering a secret, and the set from which secrets are drawn plays no role in these attacks.
		For the operad platforms in the two original papers (fields, the integers, residue rings, the tropical semiring with finite entries, convexors and the masked tropical algebras), the relevant problems can be solved in polynomial time in the input models specified here, and masking does not affect these attacks.
		The authentication protocol fails for a more elementary reason: as originally stated it does not bind the prover to any public key, and even with a registered public key the response expected by the verifier can be computed from public data.
		In the masked variant, verification requires the secret mask; on the concrete string platforms, one observed session with an invertible challenge entry suffices for forgery.
		We also observe that one of the auxiliary constructions satisfies the algebra axioms only in a degenerate case, although the corresponding protocol remains correct.
		The attacks are implemented and verified on random instances of representative platforms.
	\end{abstract}


	\section{Introduction}
	\label{sec:introduction}

	Algebraic cryptography looks for algebraic structures (\emph{platforms}) on which some computational problem is believed to be hard, typically groups~\cite{myasnikov2011noncommutative}.
	\emph{Commutative operads}, introduced by Tronin~\cite{tronin2006operads}, were proposed as such a platform in~\cite{gaynullina2016operad}.
	In a commutative operad two operations can be applied ``in either order'' (see~\eqref{eq:comm} below), and this plays the role of commutativity in Diffie--Hellman-type protocols.
	Four protocols are described in~\cite{gaynullina2016operad} (common key creation, key exchange, encryption and authentication), and the first one is analyzed on operads built from commutative semirings, with the tropical semiring indicated as a promising choice~\cite[Example~7]{gaynullina2016operad}.
	The follow-up~\cite{gaynullina2016platforms} introduces a \emph{masking} of the operad by a secret invertible element and a family of algebras with secret parameters.
	The security of the resulting tropical scheme is related there to a system of nonlinear equations in all the secret quantities.

	In this paper we analyze these protocols.
	Our main observation is that an attacker never needs to solve the problems on which their security was based.
	It suffices to find \emph{some} operation that commutes with the public ones and is consistent with the transmitted data.
	On all the operad platforms proposed in~\cite{gaynullina2016platforms,gaynullina2016operad} this is a linear problem.
	The authentication protocol admits an even more direct attack, since its verification uses only public data.
	We analyze the operad constructions of these two papers in the input models specified in Section~\ref{sec:platforms}, with finite data in the tropical case.
	The analysis is restricted to these operadic constructions: we do not treat the groupoid platform of~\cite[Section 3]{gaynullina2016platforms}, which does not involve operads, and we make no claim about efficient attacks on arbitrary entropic algebras.

	\paragraph{Contributions}
	\begin{enumerate}[label=(\roman*)]
		\item We show that an eavesdropper never needs the secret.
			For key agreement it suffices to solve an \emph{interpolation problem} $\INT$ over the centralizer of the public operations (Theorem~\ref{thm:int}).
			For key exchange and encryption it suffices to solve a \emph{division problem} $\DIV$ over the centralizer of the secret space in an ambient operad (Lemma~\ref{lem:factor} and Theorem~\ref{thm:div}).
			Any solution works.
			The subset from which secrets are drawn, including the restriction to $k$-th powers in~\cite[Theorem~2]{gaynullina2016operad}, plays no role.
		\item We show that the authentication protocol does not authenticate: as stated it accepts every prover, and with a registered public key the expected response is computable from public data (Proposition~\ref{prop:auth}).
		\item For the operad platforms in~\cite{gaynullina2016platforms,gaynullina2016operad} we solve the relevant instances of $\INT$ and $\DIV$ in polynomial time, with finite tropical data and the representation of the inputs made explicit (Section~\ref{sec:platforms}).
			The tropical case reduces to \emph{one-sided} min-plus systems, which have closed-form solutions.
			The hardness results cited in~\cite{gaynullina2016operad} concern two-sided systems.
		\item Masking by $c$ is equivalent to replacing each secret $w$ by $w[c]$, so it is transparent to the attacks on Protocols~1--3 (Proposition~\ref{prop:mask}).
			The secret exponents and offsets of the algebras of~\cite{gaynullina2016platforms} are absorbed componentwise (Theorem~\ref{thm:masked}).
			The construction in~\cite[Lemma~1]{gaynullina2016platforms} satisfies the algebra axioms only when the offsets are trivial, which does not affect the correctness of the protocol (Proposition~\ref{prop:lemma1}).
		\item We implement the attacks on representative platforms and verify them on random instances, including instances where the attacker's solution differs from the secret, and we check the permutation identities of the factorization lemma exactly (Section~\ref{sec:experiments}).
	\end{enumerate}

	\paragraph{Related work}
	The phenomenon is the operadic analogue of the linear decomposition attack of Myasnikov and Roman'kov~\cite{myasnikov2015linear}: the public data generate a structure in which the shared key is determined by solving a linear problem, whatever the secrets are.
	The tropical part parallels the analysis by Kotov and Ushakov~\cite{kotov2018analysis} of the tropical scheme of Grigoriev and Shpilrain~\cite{grigoriev2014tropical}.
	Here the centralizer provides a common framework for attacks on key agreement, key exchange and encryption, expressed as interpolation and division problems.
	For the masked construction of~\cite{gaynullina2016platforms}, the secret exponents, offsets and mask enter only through effective coefficients in a known family of componentwise linear operations.
	Recovering any compatible coefficients therefore suffices, without solving the nonlinear system in the original secret parameters.
	The framework also gives a necessary condition for security: interpolation and division must resist attacks in every efficiently representable family of commuting operations containing the honest ones.

	Algebras over commutative operads are \emph{entropic} (or medial) algebras, whose operations commute pairwise (Section~\ref{sec:operads}).
	Gligoroski~\cite{gligoroski2021entropoid} proposed entropic quasigroups (\emph{entropoids}) as a post-quantum replacement for groups in Diffie--Hellman.
	Panny~\cite{panny2021entropoids} gave a polynomial-time reduction of the discrete-logarithm problem for the concrete proposed instances to the ordinary problem in the underlying finite field.
	A classical structure theorem helps explain this result: every entropic quasigroup has the form $x \cdot y = \varphi(x) + \psi(y) + c$ for an abelian group $(G, +)$, commuting automorphisms $\varphi$ and $\psi$, and a constant $c$.
	Murdoch introduced these ``abelian quasi-groups'' and constructed them in this way from abelian groups~\cite{murdoch1939quasigroups}.
	The representation theorem was proved by Toyoda~\cite[Theorem 1]{toyoda1941axioms} and by Bruck~\cite[\S10]{bruck1944some}, who attributes it essentially to Murdoch's two papers~\cite{murdoch1939quasigroups,murdoch1941structure}.
	Toyoda also proves the $n$-ary version, $(x_1, \ldots, x_n) \mapsto \sum_k A_k x_k + c$~\cite[Theorem 3]{toyoda1941axioms}.
	Gligoroski~\cite{gligoroski2021rebuttal} accepts Panny's result for the concrete instances, but questions its extension to arbitrary entropic groupoids.
	In the revised version of~\cite[\S2.2]{panny2021entropoids}, Panny accepts that the representation need not be efficiently recoverable in general and that non-quasigroups lie outside the classical theorem, but shows that Gligoroski's concrete non-quasigroup construction succumbs to essentially the same attack.
	Thus neither paper establishes a universal efficient attack on all entropic groupoids.
	This paper is a further instance of the same broad pattern, in an operadic setting where most platforms are not quasigroups (see Section~\ref{sec:conclusions}).

	\paragraph{Outline}
	Section~\ref{sec:prelim} recalls operads, commutative operads and their algebras, fixes notation, and describes the four protocols of~\cite{gaynullina2016operad}.
	Section~\ref{sec:generic} contains the generic reductions: key agreement to an interpolation problem over the centralizer, and key exchange and encryption to a division problem; it also shows that the authentication protocol does not authenticate.
	Section~\ref{sec:platforms} solves both problems on all operad platforms of the two analyzed papers, and Section~\ref{sec:masked} treats masking and the algebras with secret parameters of~\cite{gaynullina2016platforms}.
	Section~\ref{sec:experiments} reports the computational experiments, and Section~\ref{sec:conclusions} concludes.

	\section{Preliminaries}
	\label{sec:prelim}

	\subsection{Operads, commutative operads and their algebras}
	\label{sec:operads}

	We follow the conventions of~\cite{gaynullina2016operad,tronin2006operads}; general background on operads can be found in~\cite{bernardelli2021operad,markl2002operads}.
	An operad $Z = \{Z(n)\}_{n \ge 1}$ in $(\mathsf{Set}, \times)$ consists of composition maps $\gamma \colon Z(m) \times Z(n_1) \times \cdots \times Z(n_m) \to Z(n_1 + \cdots + n_m)$, a unit $1 \in Z(1)$, and right actions of the symmetric groups $\Sigma_n$ on $Z(n)$, subject to the usual associativity, unit and equivariance axioms.
	All protocols considered here only ever substitute the same element into every input of another one, so it is convenient to write, for $f \in Z(m)$ and $g \in Z(n)$,
	\[
		\pw{f}{g} \;\coloneqq\; \gamma(f; g, \ldots, g) \in Z(mn).
	\]
	In this notation associativity gives $\pw{(\pw{f}{g})}{h} = \pw{f}{(\pw{g}{h})}$, and equivariance gives $\pw{(f\sigma)}{g} = (\pw{f}{g})\,\sigma\langle n \rangle$, where $\sigma\langle n\rangle \in \Sigma_{mn}$ is the block permutation that permutes $m$ blocks of size $n$ according to $\sigma$.
	For definiteness, on strings (as in Example~\ref{ex:strings} below) a permutation $\sigma$ moves the entry in position $r$ to position $\sigma(r)$, and permutations are composed from left to right, so that $(x\sigma)\tau = x(\sigma\tau)$; the block permutation $\sigma\langle n\rangle$ moves block $r$ to block $\sigma(r)$.
	All permutation formulas below are stated in this convention; with the opposite one, $\sigma_{m,n}$ is replaced by its inverse.

	The property that makes operads usable for Diffie--Hellman-type protocols is the following commutativity condition, introduced by Tronin.

	\begin{definition}[\cite{tronin2006operads}]
		An operad $Z$ is \emph{commutative} if for all $f \in Z(m)$, $g \in Z(n)$
		\begin{equation}
			\label{eq:comm}
			\pw{f}{g} \;=\; (\pw{g}{f})\,\sigma_{m,n},
		\end{equation}
		where $\sigma_{m,n} \in \Sigma_{mn}$ is the transposition of an $m \times n$ array: $\sigma_{m,n}(i + (j-1)m) = j + (i-1)n$.
	\end{definition}

	An \emph{algebra} over $Z$ is a set $A$ with maps $Z(n) \times A^n \to A$, $(f, a_1, \ldots, a_n) \mapsto f(a_1, \ldots, a_n)$, compatible with composition, unit and symmetric actions.
	Over a commutative operad, the definition translates into an interchange law that holds in every algebra~\cite{gaynullina2016operad,tronin2006operads}: for $f \in Z(n)$, $g \in Z(m)$ and $(a_{i,j}) \in A^{n \times m}$,
	\begin{equation}
		\label{eq:interchange}
		f\big(g(a_{1,\cdot}), \ldots, g(a_{n,\cdot})\big) \;=\; g\big(f(a_{\cdot,1}), \ldots, f(a_{\cdot,m})\big),
	\end{equation}
	where $a_{i,\cdot} = (a_{i,1}, \ldots, a_{i,m})$ and $a_{\cdot,j} = (a_{1,j}, \ldots, a_{n,j})$.
	In the language of universal algebra, \eqref{eq:interchange} says that any two operations of $A$ commute, that is, $A$ is an \emph{entropic} (or medial) algebra.
	As we shall see, the interchange law is the only property of commutative operads that the protocols need, and it is also the only one that our attacks use.

	All the operad platforms proposed in~\cite{gaynullina2016platforms,gaynullina2016operad} are built from a single family of examples, which we now recall.

	\begin{example}[The string operad of a commutative monoid~{\cite[Examples 2, 5]{gaynullina2016operad}}]
		\label{ex:strings}
		Let $(K, \cdot, 1)$ be a commutative monoid and $K^{\bullet}(n) = K^n$, with $\gamma(x; y_1, \ldots, y_m) = (x_1 y_1, \ldots, x_m y_m)$, where $x_i y_i$ multiplies each entry of the string $y_i$ by $x_i$, and with $\Sigma_n$ permuting entries.
		Then $K^\bullet$ is a commutative operad and $\pw{x}{g} = (x_i g_j)_{i,j}$.
		If $K$ is a commutative semiring, $A = K^d$ is a $K^\bullet$-algebra via $x(a_1, \ldots, a_n) = \sum_i x_i a_i$.
		For a fixed $k \ge 1$ and any suboperad $R \subseteq K^\bullet$, the formula $x(a_1, \ldots, a_n) = \sum_i x_i^k a_i$ defines the $R$-algebra used in~\cite[Section 4]{gaynullina2016operad}; here $R$ acts through the power map $x \mapsto (x_1^k, \ldots, x_n^k)$, which is an operad morphism because $K$ is commutative.
		The operad of simplices $\Delta \subseteq \reals^\bullet$ of Tronin is a suboperad whose algebras include convex sets (convexors)~\cite{tronin2006operads}.
	\end{example}

	\subsection{The protocols}
	\label{sec:protocols}

	We now recall the four protocols of~\cite[Section 3]{gaynullina2016operad} in the notation above.
	Throughout, the operad, the algebra and all data not explicitly declared secret are public.
	The first protocol uses an algebra; the other three use the operad acting on itself.

	\begin{description}[leftmargin=0pt, itemsep=4pt]
		\item[Protocol 1 (common secret key).] Public: an algebra $A$ and $a_{i,j} \in A$ ($i \le n$, $j \le m$).
			Alice's secret is $f \in Z(n)$ and Bob's is $g \in Z(m)$.
			Alice publishes $\alpha_j = f(a_{\cdot,j})$, Bob publishes $\beta_i = g(a_{i,\cdot})$, and the key is $\kappa = f(\beta_1, \ldots, \beta_n) = g(\alpha_1, \ldots, \alpha_m)$ by~\eqref{eq:interchange}.
		\item[Protocol 2 (key exchange).] Public: $\omega \in Z(n)$.
			Alice's secret is $\alpha \in Z(m)$ and Bob's is $\beta \in Z(k)$.
			They publish $A = \pw{\alpha}{\omega}$ and $B = \pw{\beta}{\omega}$.
			Alice computes $\pw{\alpha}{B}$ and Bob computes $\pw{\beta}{A}$, which agree up to a public permutation.
		\item[Protocol 3 (encryption).] Public: $g \in Z(n)$ and a hash function $h$.
			The secret key is $x \in Z(m)$ and the public key is $y = \pw{x}{g}$.
			To encrypt a message $\mu$, the sender chooses $k \in Z(d)$ and sends $c_1 = \pw{k}{g}$ and $c_2 = \mu \oplus h(\pw{k}{y})$.
			The receiver computes $\pw{k}{y}$ from $\pw{x}{c_1}$ and a public permutation.
		\item[Protocol 4 (authentication).] Alice's secret is $x \in Z(n)$.
			Bob sends $g \in Z(m)$ and Alice returns $y = \pw{x}{g}$.
			Bob sends a challenge $k \in Z(\ell)$ and Alice returns $z = \pw{(\pw{x}{k})}{g}$.
			Bob accepts if $z$ equals $\pw{k}{y}$ up to a public permutation.
	\end{description}

	The security of Protocol~1 is based in~\cite{gaynullina2016operad} on the difficulty of finding $\xi$ from $b_1, \ldots, b_s$ and $\xi(b_1, \ldots, b_s)$, and that of Protocols~2--4 on the difficulty of finding $x$ from $g$ and $\pw{x}{g}$.
	In both cases, the assumption is that an attacker must recover the secret element.
	The next section shows that this is not the case.

	\section{Generic attacks}
	\label{sec:generic}

	\subsection{Key agreement and the centralizer}

	Consider first Protocol~1 from the point of view of an eavesdropper.
	The only thing that makes the two keys agree is that Alice's operation commutes with Bob's.
	Rather than recovering Alice's secret, which is the problem the protocol is designed to make hard, an attacker can therefore look for \emph{any} operation that commutes with every operation Bob might use and is consistent with Alice's messages; such an operation reproduces the key.
	To make this precise we only need that every element $g \in Z(m)$ acts as an $m$-ary operation $g \colon A^m \to A$; the algebra axioms play no role.
	This will matter in Section~\ref{sec:masked}, where they fail.
	The natural home for such operations is the following set.

	\begin{definition}[Centralizer]
		\label{def:cen}
		Let $Z$ act on a set $A$ as above.
		The \emph{centralizer} $\mathrm{Cen}(Z, A)(n)$ is the set of all maps $\varphi \colon A^n \to A$ such that for all $m$, all $g \in Z(m)$ and all $(a_{i,j}) \in A^{n \times m}$,
		\[
			\varphi\big(g(a_{1,\cdot}), \ldots, g(a_{n,\cdot})\big) \;=\; g\big(\varphi(a_{\cdot,1}), \ldots, \varphi(a_{\cdot,m})\big).
		\]
	\end{definition}

	The centralizer contains the projections and is closed under substitution, so it is a clone, that is, an operad in $\mathsf{Set}$ with a canonical action on $A$.
	By construction it is the largest family of operations on $A$ that commute with the public ones.
	If $Z$ is commutative and $A$ is a $Z$-algebra, the interchange law~\eqref{eq:interchange} says precisely that the operation of every $f \in Z(n)$ lies in $\mathrm{Cen}(Z, A)(n)$.
	More generally, if $Z$ acts through an operad morphism $Z \to \Zt$ into a commutative operad $\Zt$ acting on $A$, then every operation of $\Zt$ lies in the centralizer.
	The attacker is free to search in any subfamily of the centralizer that contains Alice's operation, which leads to the following problem.

	\begin{definition}[Interpolation problem]
		For a family $\mathcal{C} \subseteq \mathrm{Cen}(Z, A)(n)$, the problem $\INT(\mathcal{C}, A)$ is the following: given $a_{i,j} \in A$ and $\alpha_1, \ldots, \alpha_m \in A$ such that $\alpha_j = \varphi(a_{\cdot,j})$ for some $\varphi \in \mathcal{C}$, find any $\varphi' \in \mathcal{C}$ with $\varphi'(a_{\cdot,j}) = \alpha_j$ for all $j$.
	\end{definition}

	The point is that a solution of this problem, whatever it is, already yields the key.

	\begin{theorem}[Protocol 1 reduces to $\INT$]
		\label{thm:int}
		Let $\varphi' \in \mathrm{Cen}(Z, A)(n)$ satisfy $\varphi'(a_{\cdot,j}) = \alpha_j$ for all $j$.
		Then $\varphi'(\beta_1, \ldots, \beta_n) = \kappa$.
	\end{theorem}
	\begin{proof}
		By definition of the centralizer, applied to Bob's secret $g$: $\varphi'(\beta_1, \ldots, \beta_n) = \varphi'\big(g(a_{1,\cdot}), \ldots, g(a_{n,\cdot})\big) = g\big(\varphi'(a_{\cdot,1}), \ldots, \varphi'(a_{\cdot,m})\big) = g(\alpha_1, \ldots, \alpha_m) = \kappa.$
	\end{proof}

	Commutativity of $Z$ enters only to guarantee that Alice's own operation lies in the centralizer, so that the instance has a solution.
	Since the attacker may choose any family $\mathcal{C}$ between Alice's secret space and the full centralizer, a platform can only be secure if $\INT(\mathcal{C}, A)$ is hard for \emph{every} family $\mathcal{C}$ with an efficient representation that contains the honest operations.
	This is a demanding requirement, because the centralizer is often much larger than the operad itself, as the following example shows.

	\begin{example}[The centralizer can be larger than the operad]
		\label{ex:matrices}
		Let $K$ be a commutative semiring and let $K^\bullet$ act on $A = K^d$ by $x(a_1, \ldots, a_n) = \sum_i x_i a_i$.
		For matrices $M_1, \ldots, M_n \in K^{d \times d}$, the map $\varphi(a_1, \ldots, a_n) = \sum_i M_i a_i$ lies in $\mathrm{Cen}(K^\bullet, A)(n)$, since scalars commute with matrices.
		The attacker can thus use $n d^2$ unknowns instead of $n$.
	\end{example}

	\subsection{Key exchange, encryption, authentication}

	In Protocols~2--4 there is no separate algebra: the operad acts on itself.
	The same idea applies, but the commuting elements now have to be found inside an \emph{ambient operad} $O \supseteq Z$; for the string platforms one takes $O = K^\bullet$.
	In the spirit of Tronin's center of a multicategory~\cite{tronin2011natural}, we define the relative centralizer
	\[
		\mathrm{Cen}_O(Z)(m) \;=\; \big\{\, x \in O(m) \mid \pw{x}{h} = (\pw{h}{x})\,\sigma_{m,d}
		\text{ for all } d \text{ and } h \in Z(d) \,\big\}.
	\]
	If $Z$ is commutative then $Z \subseteq \mathrm{Cen}_O(Z)$, and if $O$ itself is commutative then $\mathrm{Cen}_O(Z) = O$.
	Since all three protocols publish an element of the form $\pw{x}{g}$, the relevant problem for the attacker is now a division.

	\begin{definition}[Division problem]
		For a family $\mathcal{C} \subseteq \mathrm{Cen}_O(Z)(m)$, the problem $\DIV(\mathcal{C})$ is the following: given $g \in O(n)$ and $y \in O(mn)$ such that $y = \pw{x}{g}$ for some $x \in \mathcal{C}$, find any $x' \in \mathcal{C}$ with $\pw{x'}{g} = y$.
	\end{definition}

	The role played by the interchange law in Theorem~\ref{thm:int} is now taken by a factorization identity.
	It says that substituting an element of the form $\pw{h}{g}$ into $f$ only depends on $f$ through the public value $\pw{f}{g}$.

	\begin{lemma}[Factorization]
		\label{lem:factor}
		For $f \in \mathrm{Cen}_O(Z)(m)$, $g \in O(n)$ and $h \in Z(d)$,
		\[
			\pw{f}{(\pw{h}{g})} \;=\; \pw{h}{(\pw{f}{g})}\;\tau, \qquad \tau = \sigma_{m,d}\langle n\rangle.
		\]
		In particular, $f \mapsto \pw{f}{(\pw{h}{g})}$ depends on $f$ only through $\pw{f}{g}$.
	\end{lemma}
	\begin{proof}
		By associativity in $O$, the definition of $\mathrm{Cen}_O(Z)$, equivariance, and associativity again: $\pw{f}{(\pw{h}{g})} = \pw{(\pw{f}{h})}{g} = \pw{\big((\pw{h}{f})\sigma_{m,d}\big)}{g} = \pw{(\pw{h}{f})}{g}\;\sigma_{m,d}\langle n \rangle = \pw{h}{(\pw{f}{g})}\;\sigma_{m,d}\langle n \rangle$.
	\end{proof}

	The identity of Lemma~\ref{lem:factor}, with the explicit permutation $\tau$ in the convention of Section~\ref{sec:operads}, is also checked exactly on random instances in Section~\ref{sec:experiments}.

	In each of the three protocols, the quantity that the attacker wants is obtained by substituting an element of the form $\pw{h}{g}$, with $h$ chosen by the other party, into the secret.
	The lemma therefore applies directly.

	\begin{theorem}[Protocols 2--4 reduce to $\DIV$]
		\label{thm:div}
		Let $x'$ and $\alpha'$ be elements of $\mathrm{Cen}_O(Z)$ with $\pw{x'}{g} = \pw{x}{g}$ and $\pw{\alpha'}{\omega} = \pw{\alpha}{\omega}$.
		Then:
		\begin{enumerate}[label=(\alph*)]
			\item (Protocol 2) $\pw{\alpha'}{B} = \pw{\alpha}{B}$, that is, the attacker obtains Alice's key;
			\item (Protocol 3) $\pw{x'}{c_1} = \pw{x}{c_1}$, that is, the attacker decrypts every ciphertext;
			\item (Protocol 4 with a registered public key, as in Proposition~\ref{prop:auth}(b)) $\pw{(\pw{x'}{k})}{g} = \pw{(\pw{x}{k})}{g}$, that is, the attacker answers every challenge.
		\end{enumerate}
	\end{theorem}
	\begin{proof}
		(a) Since $B = \pw{\beta}{\omega}$ with $\beta \in Z$, Lemma~\ref{lem:factor} gives $\pw{\alpha'}{B} = \pw{\beta}{(\pw{\alpha'}{\omega})}\,\tau = \pw{\beta}{A}\,\tau = \pw{\alpha}{B}$.
		(b) Likewise with $c_1 = \pw{k}{g}$, $k \in Z$.
		(c) By associativity $\pw{(\pw{x}{k})}{g} = \pw{x}{(\pw{k}{g})}$, and Lemma~\ref{lem:factor} applies.
	\end{proof}

	As for key agreement, the solution found by the attacker may differ from the secret, and in the experiments of Section~\ref{sec:experiments} it usually does.
	The hardness of recovering the secret, which is the assumption made in~\cite{gaynullina2016operad}, is therefore not the relevant quantity.
	What matters is the hardness of $\INT$ and $\DIV$ over every family of operations in the centralizer that the attacker can represent efficiently.

	\subsection{The authentication protocol}

	Part~(c) of Theorem~\ref{thm:div} concerns a version of Protocol~4 in which $y$ is a registered public key.
	Protocol~4 has, however, a structural weakness that makes the division problem unnecessary.
	In its original specification, the element $y$ is sent by the prover during the session, and the verifier has no independent information about Alice against which to compare it, so the verification only checks the internal consistency of the transcript.
	Registering $y$ as Alice's public key associates it with an identity, but the verifier's decision rule still depends only on publicly computable quantities: a response is accepted if it equals $\pw{k}{y}$ up to a public permutation, and this value involves only the public key and the challenge.

	\begin{proposition}[Protocol 4 does not authenticate]
		\label{prop:auth}
		Let $x \in Z(n)$, $g \in Z(m)$, $k \in Z(\ell)$, and $\tau = \sigma_{n,\ell}\langle m\rangle$.
		\begin{enumerate}[label=(\alph*)]
			\item In Protocol~4 as stated in~\cite{gaynullina2016operad}, the verifier holds no data tied to Alice.
				For every $x'' \in Z(n)$, the responses $y'' = \pw{x''}{g}$ and $z'' = \pw{(\pw{x''}{k})}{g}$ pass verification.
			\item Suppose that $g$ is fixed and public, and that $y = \pw{x}{g}$ is Alice's registered public key, so that the verifier accepts $z$ if and only if $z = \pw{k}{y}\,\tau$.
				Then for every challenge $k$, the response $z = \pw{k}{y}\,\tau$, computed from public data only, is accepted.
				It coincides with Alice's honest response $\pw{(\pw{x}{k})}{g}$.
		\end{enumerate}
	\end{proposition}
	\begin{proof}
		By associativity and Lemma~\ref{lem:factor}, $\pw{(\pw{w}{k})}{g} = \pw{w}{(\pw{k}{g})} = \pw{k}{(\pw{w}{g})}\,\tau$ for every $w \in Z(n)$.
		For $w = x''$ this is the verification condition of~(a).
		For $w = x$ it shows that the honest response equals $\pw{k}{y}\,\tau$, and~(b) follows, since $k$, $y$ and $\tau$ are public.
	\end{proof}

	Theorem~\ref{thm:div}(c) therefore describes a valid but unnecessarily indirect attack.
	A meaningful authentication protocol needs a response that cannot be computed from public data alone.
	Within the design of~\cite{gaynullina2016operad}, the masking of~\cite{gaynullina2016platforms} makes the verification formula depend on a secret held by the verifier as well.
	Section~\ref{sec:masked} analyzes this variant and its remaining vulnerabilities.

	\section{The proposed platforms}
	\label{sec:platforms}

	We now solve the relevant interpolation and division problems on the operad platforms of~\cite{gaynullina2016platforms,gaynullina2016operad}, with finite data in the tropical case.
	Complexity statements refer to two models.
	Over an abstract field or over $\reals$, we count arithmetic operations.
	For finitely encoded inputs (elements of finite fields and residue rings, integers and rationals in binary), \emph{polynomial time} means polynomial in the bit size of the input.
	All the platforms are built from the string operad $K^\bullet$ of Example~\ref{ex:strings}.
	Since $K^\bullet$ is commutative, the attacker can work in the whole of it, and the division problem becomes elementary.

	\begin{proposition}[$\DIV$ on string operads]
		\label{prop:div}
		Let $K$ be a commutative monoid, $Z \subseteq O = K^\bullet$, and take $\mathcal{C} = O(m)$, which equals $\mathrm{Cen}_O(Z)(m)$ because $O$ is commutative.
		Then $\pw{x'}{g} = y$ means $x'_i g_j = y_{i,j}$ for all $i$ and $j$, so $\DIV$ decouples into $m$ problems in a single unknown.
		In particular:
		\begin{enumerate}[label=(\alph*)]
			\item if some $g_j$ is invertible (always in a group; any $g_j \ne 0$ in a field; any finite $g_j$ in the tropical semiring), then $x'_i = y_{i,j} g_j^{-1}$ recovers the secret, with one inversion and $m$ multiplications;
			\item for $K = \integers$ (multiplicative monoid), work in $\Q^\bullet \supseteq \integers^\bullet$: $x'_i = y_{i,j}/g_j$ for any $g_j \ne 0$;
			\item for $K = \integers/N$, each coordinate is a system of linear congruences $x'_i g_j \equiv y_{i,j} \pmod N$, which can be solved in time polynomial in $\log N$ and the number of congruences~\cite{schrijver1986theory};
			\item for the simplex operad $\Delta$ with rational entries, case~(a) applies in $\Q^\bullet$.
		\end{enumerate}
	\end{proposition}
	\begin{proof}
		Since $O$ is commutative, $\mathrm{Cen}_O(Z) = O$, and by Example~\ref{ex:strings} $\pw{x'}{g} = (x'_i g_j)_{i,j}$.
		Hence $\pw{x'}{g} = y$ holds if and only if $x'_i g_j = y_{i,j}$ for all $i$ and $j$, which for each $i$ is a system in the single unknown $x'_i$, independent of the other coordinates.
		The secret $x$ is a solution.
		(a) If $g_j$ is invertible, every solution satisfies $x'_i = y_{i,j} g_j^{-1}$, so the solution is unique and equals $x$.
		(b) The operad $\integers^\bullet$ is contained in the commutative operad $\Q^\bullet$; $x$ is still a solution in $\Q^\bullet$, and every $g_j \ne 0$ is invertible in $\Q$, so~(a) applies with $O = \Q^\bullet$.
		(c) For fixed $i$, the congruences $x'_i g_j \equiv y_{i,j} \pmod N$ are the linear Diophantine system $x'_i g_j + N t_j = y_{i,j}$ in the unknowns $x'_i$ and $t_1, \ldots, t_n$, which can be solved in polynomial time via the Hermite normal form~\cite{schrijver1986theory}.
		(d) For rational entries, $\Delta \subseteq \Q^\bullet$, and the entries of $g \in \Delta(n)$ sum to $1$, so some $g_j > 0$ is invertible in $\Q$; apply~(a) with $O = \Q^\bullet$.
	\end{proof}

	If $g$ is the zero string over a field or over $\integers$, then $y$ is also zero and any $x'$ of the required arity solves the division problem.

	For key agreement, the algebras of~\cite{gaynullina2016operad} are free modules over a semiring, and the action of a secret is a linear combination of the inputs with coefficients that are $k$-th powers.
	Forgetting the restriction to $k$-th powers, the interpolation problem becomes a linear system.

	\begin{corollary}[$\INT$ on the algebras of~\cite{gaynullina2016operad}]
		\label{cor:int}
		Let $K$ be a commutative semiring, $A = K^d$, $R \subseteq K^\bullet$ a suboperad, and $x(a_1, \ldots, a_n) = \sum_i x_i^k a_i$ for $x \in R(n)$.
		Here $R$ acts through the power map $x \mapsto (x_1^k, \ldots, x_n^k)$, which is an operad morphism $R \to K^\bullet$ because $K$ is commutative.
		Hence the maps $u(a_1, \ldots, a_n) = \sum_i u_i a_i$, $u \in K^n$, form a family $\mathcal{C} \subseteq \mathrm{Cen}(R, A)(n)$ that contains the honest operations.
		$\INT(\mathcal{C}, A)$ is the one-sided linear system $\sum_i u_i a_{i,j} = \alpha_j$ ($md$ scalar equations in $n$ unknowns).
		Hence:
		\begin{enumerate}[label=(\alph*)]
			\item for a field, Gaussian elimination solves it with $O(md \cdot n \cdot \min(md, n))$ field operations; over a finite field this is polynomial in the bit size;
			\item for $K = \integers$, solve over $\Q$ (the operad $\Q^\bullet$ acts on $\Q^d \supseteq \integers^d$); Gaussian elimination over $\Q$ runs in time polynomial in the bit size of the input~\cite{schrijver1986theory}, and the key obtained from a rational solution is the integer key;
			\item for the tropical semiring $(\reals \cup \{+\infty\}, \min, +)$, assuming that all entries $a_{i,j}[t]$ and all messages $\alpha_j[t]$ are finite, the principal solution $\bar u_i = \max_{j,t}\big(\alpha_j[t] - a_{i,j}[t]\big)$ solves it with $O(nmd)$ additions and comparisons of input numbers~\cite{butkovic2010maxlinear,cuninghame1979minimax};
			\item for $K = \integers/N$, it is a linear system over $\integers/N$, which reduces to a system of linear Diophantine equations and is solvable in polynomial time~\cite{schrijver1986theory}.
		\end{enumerate}
	\end{corollary}
	\begin{proof}
		A solution exists (the vector of $k$-th powers of the secret), so the only point is~(c).
		Let $u^\star$ be a solution, possibly with infinite coordinates.
		Since the public entries and messages are finite, $\bar u$ is finite.
		From $u^\star_i + a_{i,j}[t] \ge \alpha_j[t]$ for all $i$, $j$ and $t$ we get $\bar u \le u^\star$, hence $\min_i(\bar u_i + a_{i,j}[t]) \le \alpha_j[t]$.
		The reverse inequality holds by definition of $\bar u$.
	\end{proof}

	Case~(c) deserves a comment, since the tropical semiring was singled out in~\cite[Example~7]{gaynullina2016operad} as a promising choice, with reference to results on the complexity of tropical linear systems.
	Those results concern \emph{two-sided} systems, with unknowns on both sides of the equations.
	The system arising here is \emph{one-sided}, and its principal solution is given by a closed formula.
	The finiteness assumption covers the scheme of~\cite{gaynullina2016platforms}, which works over $\integers$.
	Entries equal to $+\infty$ require the usual conventions of max-linear algebra, including equations whose entries are all infinite and variables that occur in no finite term, and we do not treat them here.

	The convexors of~\cite{tronin2006operads} need a separate statement, because they are not modules over a semiring.
	We consider them in the coordinate model in which they arise as examples: $A$ is a convex subset of $\reals^d$, the public data have rational coordinates, and $\lambda \in \Delta(n)$ acts by $\lambda(a_1, \ldots, a_n) = \sum_i \lambda_i a_i$.
	Since $\Delta$ is commutative, the maps $a \mapsto \sum_i u_i a_i$ with $u \in \Delta(n)$ lie in $\mathrm{Cen}(\Delta, A)(n)$.
	The interpolation problem over this family is the linear feasibility problem
	\[
		u_i \ge 0, \qquad \sum_i u_i = 1, \qquad \sum_i u_i\, a_{i,j} = \alpha_j \quad (j = 1, \ldots, m),
	\]
	which can be solved in time polynomial in the bit size of the data by linear programming~\cite{schrijver1986theory}.
	We do not claim anything for abstract convex algebras that come without such a linear coordinate representation.

	\section{Masking and secret parameters}
	\label{sec:masked}

	The follow-up paper~\cite{gaynullina2016platforms} proposes two ways of strengthening the protocols: masking the operad by a secret element, and using algebras whose structure depends on secret parameters.
	We treat them in turn.

	\subsection{Masking}

	For an invertible $c \in Z(1)$, \cite[Section 4]{gaynullina2016platforms} defines the masked operad $Z_c$, with composition $[w; w_1, \ldots, w_m]_c = \gamma(w; \gamma(c; w_1), \ldots, \gamma(c; w_m))$, and the masked algebra $A_c$, with $[w\, a_1 \ldots a_n]_c = (\pw{w}{c})(a_1, \ldots, a_n)$.
	The intent is that, once $c$ is shared secretly, the platform itself becomes unknown to the attacker.
	A short computation shows that this is not the case: the mask can always be moved onto the secrets.

	\begin{proposition}[Masking is transparent]
		\label{prop:mask}
		Let $Z$ be a commutative operad, $A$ a $Z$-algebra, and $c \in Z(1)$ invertible.
		Then:
		\begin{enumerate}[label=(\alph*)]
			\item $[w; w_1, \ldots, w_m]_c = \gamma(\pw{w}{c}; w_1, \ldots, w_m)$;
			\item every run of Protocols 1--3 over $(Z_c, A_c)$ produces the same public data and keys as a run over $(Z, A)$ in which each secret $w$ is replaced by $\pw{w}{c} \in Z$;
			\item masking $Z_{c_1}$ by $c_2$ gives $Z_{c'}$ with $c' = \gamma(c_1; \gamma(c_2; c_1)) \in Z(1)$.
		\end{enumerate}
		Consequently Theorems~\ref{thm:int} and~\ref{thm:div}(a)--(b) apply verbatim to masked Protocols~1--3, and the attacker never needs $c$.
	\end{proposition}
	\begin{proof}
		(a) is associativity: $\gamma(w; \gamma(c; w_1), \ldots) = \gamma(\gamma(w; c, \ldots, c); w_1, \ldots)$.
		(b) follows from (a) and the definition of $A_c$, since in Protocols~1--3 every masked composition has a secret element (or a value computed from one) in first position.
		(c) By (a), the masked composition of $Z_{c_1}$ by $c_2$ is $\gamma(\pw{w}{c_1}; \gamma(\pw{c_2}{c_1}; w_1), \ldots) = \gamma(\pw{(\pw{w}{c_1})}{(\pw{c_2}{c_1})}; w_1, \ldots)$, and $\pw{(\pw{w}{c_1})}{(\pw{c_2}{c_1})} = \pw{w}{c'}$ by associativity.
	\end{proof}

	The iterated masking procedure of~\cite[Section 2]{gaynullina2016platforms} is meant to make the attacker ``lag behind'' by a constant time per round.
	By parts (b) and (c), iterating the masking leaves the attacks on Protocols~1--3 unchanged.

	Protocol~4 behaves differently, because the verifier also has to compute with the mask.
	With a registered public key as in Proposition~\ref{prop:auth}(b), the public key is $y = [x; g, \ldots, g]_c = \pw{(\pw{x}{c})}{g}$.
	By Proposition~\ref{prop:mask}(a), associativity and Lemma~\ref{lem:factor}, the expected response to a challenge $k$ is $z = \pw{(\pw{x}{c})}{(\pw{(\pw{k}{c})}{g})} = \pw{(\pw{k}{c})}{y}\,\tau$.
	Hence the verifier must know $c$, and verification requires a shared secret.
	The expected response now depends on that secret, so neither the public forgery of Proposition~\ref{prop:auth}(b) nor Theorem~\ref{thm:div}(c) applies directly: the attacker cannot form $\pw{k}{c}$.
	On string operads, however, a single observed session with an invertible challenge entry reveals enough about the mask.

	\begin{proposition}[One transcript suffices on string operads]
		\label{prop:masked4}
		Let $Z \subseteq K^\bullet$ with $K$ a commutative monoid, and suppose the attacker observes one accepted transcript $(k, z)$ for the public key $y$, in which some entry $k_l$ of $k$ is invertible.
		Let $c' \in K$ be any solution of $k_l c' y_r = k_l c\, y_r$ for all $r$; the right-hand sides are entries of $z$, up to the public permutation $\tau$.
		Then for every later challenge $k'$, the response $\pw{(\pw{k'}{c'})}{y}\,\tau$ is accepted.
	\end{proposition}
	\begin{proof}
		Since $k_l$ is invertible, $c' y_r = c\, y_r$ for all $r$.
		The accepted response $\pw{(\pw{k'}{c})}{y}\,\tau$ has entries $k'_{l'} c\, y_r$, and these equal $k'_{l'} c'\, y_r$.
	\end{proof}

	Finding $c'$ is a division problem in a single unknown.
	On the concrete platforms of Proposition~\ref{prop:div}, it is solvable in polynomial time.
	The attacker need not recover $x$ or $c$: it suffices to find $c'$ with $c'y_r = cy_r$ for all $r$.
	Over a commutative ring, these possible masks form the coset of $c$ modulo the common annihilator of the entries of $y$.

	\subsection{Algebras with secret parameters}

	The second strengthening concerns key agreement.
	For a commutative semiring $K$, positive integers $k_1, \ldots, k_d$ and $y_1, \ldots, y_d \in K$, \cite[Section 4]{gaynullina2016platforms} defines on $A = K^d$ the operations
	\begin{equation}
		\label{eq:Aky}
		x(a_1, \ldots, a_n)_t \;=\; \sum_{i} a_{i}[t]\, x_i^{k_t}\, y_t, \qquad t = 1, \ldots, d,
	\end{equation}
	and, after masking by $c$ and in the tropical case, takes $x$, the $k_t$, the $y_t$ and $c$ to be secret.
	Security is then related in~\cite[Theorem 4]{gaynullina2016platforms} to the system $\min_i \big(a_{i,j}[t] + k_t c + k_t x_i + y_t\big) = \alpha_j[t]$ (written here in the notation of~\eqref{eq:Aky}), which is nonlinear in all these unknowns.
	Before turning to the attack, we observe that~\eqref{eq:Aky} is an algebra only in a degenerate case.

	\begin{proposition}[On {\cite[Lemma 1]{gaynullina2016platforms}}]
		\label{prop:lemma1}
		Let $R \subseteq K^\bullet$ be a suboperad (so $1 \in R(1)$).
		Then~\eqref{eq:Aky} defines an $R$-algebra if and only if every $y_t$ is the multiplicative identity $1$ of $K$ (which is $0$ in the tropical semiring).
		Nevertheless, Protocol~1 with~\eqref{eq:Aky} is correct for arbitrary $y_t$.
	\end{proposition}
	\begin{proof}
		The unit acts as $1(a)_t = a[t]\, y_t$.
		Taking $a[t] = 1$, the unit axiom forces $y_t = 1$.
		For $y_t = 1$ the structure is that of Example~\ref{ex:strings} with componentwise exponents, and it is an algebra.
		For correctness, both $f(\beta_1, \ldots, \beta_n)_t$ and $g(\alpha_1, \ldots, \alpha_m)_t$ equal $\sum_{i,j} a_{i,j}[t]\, f_i^{k_t} g_j^{k_t} y_t^2$.
	\end{proof}

	Associativity alone is weaker: nested actions carry $y_t^2$ while the composite carries $y_t$, so it forces $y_t^2 = y_t$ (apply associativity to two unary operad units and an input equal to $1$).
	Over a general semiring this also allows, for instance, $y_t = 0$; for finite tropical offsets it again means that $y_t$ is the multiplicative identity.
	The reduction in~\cite[Theorem 4]{gaynullina2016platforms} relies on~\cite[Theorem 2]{gaynullina2016operad}, which is stated for algebras, so it applies only in this case.
	This has no consequence for the attack, which never uses the algebra axioms.
	The secret parameters simply disappear into the unknowns of a linear system.

	\begin{theorem}[Secret parameters are absorbed]
		\label{thm:masked}
		Consider Protocol~1 with~\eqref{eq:Aky}, masked by $c$, with $x$, $(k_t)$, $(y_t)$ and $c$ secret.
		For each $t$, the vector $u^{(t)}_i = (c\,x_i)^{k_t} y_t$ satisfies the one-sided linear system $\sum_i a_{i,j}[t]\, u^{(t)}_i = \alpha_j[t]$, $j = 1, \ldots, m$.
		For \emph{any} solution $\bar u^{(t)}$ of this system, $\kappa_t = \sum_i \beta_i[t]\, \bar u^{(t)}_i$.
	\end{theorem}
	\begin{proof}
		By definition of the masked action, $[x\, a_1 \ldots a_n]_c = (\pw{x}{c})(a_1, \ldots, a_n)$, and $\pw{x}{c} = (c x_1, \ldots, c x_n)$ in the string operad.
		Hence the masked action is~\eqref{eq:Aky} with $x_i$ replaced by $c x_i$, so $\alpha_j[t] = \sum_i a_{i,j}[t]\, u^{(t)}_i$ and, with Bob's analogous vector $v^{(t)}$, $\beta_i[t] = \sum_j a_{i,j}[t]\, v^{(t)}_j$ and $\kappa_t = \sum_j v^{(t)}_j \alpha_j[t]$.
		Then $\sum_i \beta_i[t]\, \bar u^{(t)}_i = \sum_j v^{(t)}_j \sum_i a_{i,j}[t]\, \bar u^{(t)}_i = \sum_j v^{(t)}_j \alpha_j[t] = \kappa_t$.
	\end{proof}

	The attacker therefore solves $d$ independent one-sided systems as in Corollary~\ref{cor:int}.
	In the tropical case with finite data, which is the setting of~\cite{gaynullina2016platforms} over $\integers$, this takes $O(nmd)$ operations, and $x$, $k_t$, $y_t$ and $c$ are never computed.
	Theorem~\ref{thm:masked} is in fact an instance of Theorem~\ref{thm:int}: every operation $a \mapsto \big(\sum_i a_i[t]\, u^{(t)}_i\big)_t$ with $u^{(t)} \in K^n$ commutes with every masked operation of Bob, which has the same componentwise form, so these maps lie in the centralizer of the public action.
	This holds whether or not~\eqref{eq:Aky} is an algebra, which is why Proposition~\ref{prop:lemma1} does not affect the attack.
	The centralizer formulation also avoids requiring the unmasked coefficient map $x \mapsto (x_i^{k_t} y_t)_{i,t}$ to be an operad morphism: when some $y_t \ne 1$, this map does not preserve the unit.

	\section{Computational experiments}
	\label{sec:experiments}

	Since the results of Sections~\ref{sec:generic}--\ref{sec:masked} are proved, the experiments are not needed to establish them.
	They serve two other purposes: to check the main statements, including the explicit permutations, on random instances of representative platforms, and to show that the attacks are practical even with a straightforward implementation.

	\subsection{Setup}

	The protocols and attacks in the experiments are implemented in Python~3.14 using only the standard library.
	Exact rational arithmetic is provided by the \texttt{fractions} module, and modular arithmetic is native.
	The experiments run on a single thread of a laptop with an Intel Core i7-1185G7 processor (3.0\,GHz) and 16\,GB of RAM.

	The experiments cover representative platforms rather than every case of Section~\ref{sec:platforms}:
	\begin{itemize}
		\item Protocol~1 over $\mathrm{GF}(2^{61}-1)$, over $\integers$ (solved over $\Q$) and over the tropical semiring with finite entries, together with its masked variant with secret exponents and offsets;
		\item Protocols~2 and~3 over $\integers/2^{12}$;
		\item Protocol~4 over $\integers/2^{12}$ in three versions: as stated, with a registered public key, and masked.
	\end{itemize}
	The interpolation problem for convexors and over residue rings is not implemented; for these cases we rely on the proofs and on the standard algorithms cited in Section~\ref{sec:platforms}.
	In addition, for random elements of the string operad over $\integers/2^{12}$, we check exactly the commutativity identity~\eqref{eq:comm}, equivariance, and the identity of Lemma~\ref{lem:factor}, with the explicit permutations $\sigma_{m,d}$, $\sigma\langle n\rangle$ and $\sigma_{m,d}\langle n\rangle$ in the convention of Section~\ref{sec:operads}.
	These checks fail if $\sigma_{m,d}$ is replaced by its inverse.
	The same explicit permutations are used whenever a protocol compares values up to a public permutation.

	For every configuration we run 10 independent trials, with random seeds $0, \ldots, 9$.
	Each trial draws the public data and all secrets uniformly at random, as follows.
	\begin{itemize}
		\item Field $\mathrm{GF}(p)$, $p = 2^{61}-1$: entries in $\{1, \ldots, p-1\}$.
		\item Integers: entries in $\{1, \ldots, 50\}$, to keep the rational arithmetic small.
		\item Tropical semiring: integer entries in $[-10^6, 10^6]$.
			In the masked variant the exponents $k_t$ are drawn from $\{1, \ldots, 10^6\}$, and the offsets $y_t$ and the mask $c$ from $[-10^6, 10^6]$.
		\item Residue ring $\integers/2^{12}$ (Protocols~2--4): secrets are uniform, while the public elements $g$ and $\omega$ have even entries.
			Then division is never unique, so the experiments test the claim that \emph{any} solution works.
			In the masked Protocol~4 the mask $c$ is odd, hence invertible, and one entry of the observed challenge is odd, as required by Proposition~\ref{prop:masked4}.
	\end{itemize}
	The sizes $(n, m, d, k)$ follow the notation of Sections~\ref{sec:protocols} and~\ref{sec:platforms}, and include both over- and underdetermined interpolation problems.

	The attacker only receives the public data and the transmitted messages.
	For Protocol~1 over a field or over $\integers$, it solves the linear system of Corollary~\ref{cor:int} by Gaussian elimination, working over $\Q$ in the integer case and setting free variables to zero.
	In the tropical case, with or without masking, it uses the principal solution of Corollary~\ref{cor:int}(c), componentwise as in Theorem~\ref{thm:masked}.
	For Protocols~2--4 over $\integers/2^{12}$, it solves the division problem coordinatewise as in Proposition~\ref{prop:div}.
	For simplicity, this is done by exhaustive search over the $2^{12}$ residues, although in general one would solve a system of linear congruences.

	A trial counts as a success if the attacker's output is exactly equal to the honest value that the attack targets:
	\begin{itemize}
		\item the shared key (Protocols~1 and~2);
		\item the receiver's decryption value (Protocol~3);
		\item for Protocol~4 with a registered public key, the response accepted by the verifier, computed either through the division problem (Theorem~\ref{thm:div}(c)) or from public data only (Proposition~\ref{prop:auth}(b));
		\item for masked Protocol~4, the accepted response to a fresh challenge after one observed transcript (Proposition~\ref{prop:masked4});
		\item acceptance by the verifier of an impostor using her own secret (Protocol~4 as stated, Proposition~\ref{prop:auth}(a)).
	\end{itemize}
	We also record whether the attacker's solution coincides with the corresponding secret quantity: the vector of $k$-th powers (or $u^{(t)}$ in the masked case) for Protocol~1, the secret element for Protocols~2--4, and the mask for the masked Protocol~4.
	Times are wall-clock medians.
	For Protocol~1 they cover only the attacker's computation.
	For Protocols~2--4 they also include the simulation of the honest parties, and they are negligible in any case.
	Times should be read as indicative: on this machine they vary by up to a factor of about three between runs, depending on power management, so only orders of magnitude and growth rates are meaningful.

	\subsection{Correctness}

	The outcome is reported in Table~\ref{tab:experiments}.
	All 260 trials succeed, and the column ``Solution $=$ secret'' illustrates that recovering the secret is never needed.
	For Protocol~1 over a field or over $\integers$, the attacker finds the secret in the configurations with a unique interpolation solution, and in none of the trials with the underdetermined configuration $(60, 4, 3, 2)$.
	In the tropical case the principal solution is the \emph{least} solution and differs from the secret in every recorded trial.
	Over $\integers/2^{12}$ the division problem has many solutions, and the attacker's solution matches the secret only occasionally.

	\begin{table}[!tbh]
		\centering
		\caption{Correctness experiments, with 10 random trials per configuration. ``Solution $=$
		secret'' compares the attacker's interpolating or dividing element with the corresponding secret
		quantity; for masked Protocol~4 it compares the recovered mask with the true one. It is not
		applicable to the exact identity checks, to the public forgery and to the impostor, which do not
		compute any substitute for a secret.}
		\label{tab:experiments}
		\begin{adjustbox}{max width=\textwidth}
		\begin{tabular}{lllccr}
			\toprule
			Protocol & Platform & Sizes & Success & Solution $=$ secret & Median time (ms) \\
			\midrule
			1 & $\mathrm{GF}(2^{61}-1)$ & $(n,m,d,k) = (5,5,5,3)$ & 10/10 & 10/10 & 0.5 \\
			  &                        & $(40,30,2,7)$ & 10/10 & 10/10 & 35.9 \\
			  &                        & $(60,4,3,2)$ & 10/10 & 0/10 & 3.8 \\
			1 & $\integers$ (via $\Q$) & $(5,5,5,3)$ & 10/10 & 10/10 & 1.5 \\
			  &                        & $(40,30,2,7)$ & 10/10 & 10/10 & 299.3 \\
			  &                        & $(60,4,3,2)$ & 10/10 & 0/10 & 22.9 \\
			1 & tropical & $(5,5,5,3)$ & 10/10 & 0/10 & $<0.1$ \\
			  &          & $(40,30,2,7)$ & 10/10 & 0/10 & 0.3 \\
			  &          & $(60,4,3,2)$ & 10/10 & 0/10 & 0.2 \\
			1 masked & tropical, $k_t \le 10^6$ & $(n,m,d) = (5,5,5)$ & 10/10 & 0/10 & $<0.1$ \\
			  &          & $(40,30,8)$ & 10/10 & 0/10 & 1.5 \\
			  &          & $(60,4,3)$ & 10/10 & 0/10 & 0.2 \\
			\midrule
			Lemma~\ref{lem:factor} & $\integers/2^{12}$ & $(m,d,n) = (3,4,5)$ & 10/10 & n/a & $<0.1$ \\
			  &                   & $(6,3,2)$ & 10/10 & n/a & $<0.1$ \\
			2 & $\integers/2^{12}$ & $(n,m,k) = (3,4,5)$ & 10/10 & 1/10 & 1.9 \\
			  &                   & $(6,3,2)$ & 10/10 & 2/10 & 1.8 \\
			3 & $\integers/2^{12}$ & $(n,m,d) = (3,4,5)$ & 10/10 & 1/10 & 2.0 \\
			  &                   & $(6,3,2)$ & 10/10 & 2/10 & 2.8 \\
			4 public key, via $\DIV$ & $\integers/2^{12}$ & $(n,m,\ell) = (3,4,5)$ & 10/10 & 1/10 & 1.7 \\
			  &                   & $(6,3,2)$ & 10/10 & 0/10 & 3.0 \\
			4 public key, public forgery & $\integers/2^{12}$ & $(3,4,5)$ & 10/10 & n/a & $<0.1$ \\
			  &                   & $(6,3,2)$ & 10/10 & n/a & $<0.1$ \\
			4 as stated & $\integers/2^{12}$ & $(3,4,5)$ & 10/10 & n/a & $<0.1$ \\
			  &                   & $(6,3,2)$ & 10/10 & n/a & $<0.1$ \\
			4 masked & $\integers/2^{12}$ & $(3,4,5)$ & 10/10 & 5/10 & 10.0 \\
			  &                   & $(6,3,2)$ & 10/10 & 4/10 & 5.1 \\
			\bottomrule
		\end{tabular}
		\end{adjustbox}
	\end{table}

	\subsection{Scaling}

	To see how the attacks behave on larger instances, Table~\ref{tab:scaling} reports attacker times for Protocol~1, with 3 trials per size, all successful.
	Over $\mathrm{GF}(2^{61}-1)$ we take $m = n/2$ and $d = 2$, rounding $m$ down when $n$ is odd, so that the interpolation system is approximately square.
	Gaussian elimination is cubic, and indeed the time grows by a factor of about $8$ when $n$ doubles.
	For the masked tropical algebras we take $n = m$ and $d = 8$.
	The attack costs $O(nmd)$, and the time grows by a factor of about $4$.
	This is the same order as the cost of computing Alice's messages $\alpha_j$ in the first place.
	The attacker therefore does not lag behind the honest parties in any meaningful sense, contrary to the assumption underlying the masking procedure of~\cite[Section 2]{gaynullina2016platforms}.

	\begin{table}[!tbh]
		\centering
		\caption{Scaling of the attack on Protocol~1, with 3 trials per size, all successful.}
		\label{tab:scaling}
		\begin{tabular}{lrrrr}
			\toprule
			& \multicolumn{4}{c}{Median attacker time (s)} \\
			\cmidrule(lr){2-5}
			$\mathrm{GF}(2^{61}-1)$, $m = n/2$, $d = 2$, $k = 5$ & $n = 25$ & $n = 50$ & $n = 100$ & $n = 200$ \\
			& 0.01 & 0.04 & 0.37 & 2.65 \\
			\midrule
			masked tropical, $n = m$, $d = 8$, $k_t \le 10^6$ & $n = 100$ & $n = 200$ & $n = 400$ & $n = 800$ \\
			& 0.01 & 0.05 & 0.21 & 0.84 \\
			\bottomrule
		\end{tabular}
	\end{table}

	\paragraph{Code availability}
	The implementation of the protocols, of the attacks and of the experiment driver is publicly available at \url{https://github.com/AmbrogioMB/operad-cryptography}~\cite{bernardelli2026code}, together with the raw results reported in Tables~\ref{tab:experiments} and~\ref{tab:scaling}.


	\section{Conclusions}
	\label{sec:conclusions}

	Commutative operads were proposed as a platform for public-key cryptography in~\cite{gaynullina2016operad}.
	The proposal includes protocols for key agreement, key exchange, encryption and authentication, whose security was related to the recovery of a secret operad element from its action on public data, with the tropical semiring indicated as a promising instance.
	Masking and algebras with secret parameters were later added to strengthen the protocols~\cite{gaynullina2016platforms}.

	Our analysis shows that these protocols rely only on the interchange law, and that the same law is all an attacker needs.
	Key agreement reduces to an interpolation problem over the centralizer of the public operations, and key exchange and encryption reduce to a division problem.
	In both cases any solution suffices, and the secret itself is never required.
	On the operad platforms in~\cite{gaynullina2016platforms,gaynullina2016operad}, the relevant problems are linear and can be solved in polynomial time under the input assumptions of Section~\ref{sec:platforms}.
	In the tropical case with finite entries they have closed-form solutions.
	Masking does not affect these attacks, and the secret exponents and offsets of the parametrized algebras are absorbed componentwise into the unknowns of a linear system.
	The authentication protocol fails for a more elementary reason.
	In its original formulation it does not bind the prover to a public key at all, and with a registered public key the expected response is computable from public data.
	In the masked version, verification requires the secret mask; on the concrete string platforms, one observed session with an invertible challenge entry suffices for forgery.
	Along the way we have also observed that the parametrized algebras satisfy the algebra axioms only in a degenerate case, without consequences for the correctness of the protocol.
	The experiments confirm the main statements on random instances of representative platforms.
	Even a straightforward implementation breaks the masked tropical scheme with $n = m = 800$ in at most a few seconds on a laptop.

	That all the analyzed instances turned out to be linear is probably not a coincidence.
	Algebras over commutative operads are entropic (Section~\ref{sec:operads}).
	Over a field, an operation $\sum_i x_i a_i$ with all $x_i \ne 0$ is an $n$-ary quasigroup, and Toyoda's $n$-ary theorem~\cite{toyoda1941axioms} predicts its affine form.
	The integer, residue-ring and tropical platforms are not quasigroups, since translations need not be bijective there, so the classical structure theory does not apply to them directly.
	This is the same limitation discussed in the exchange between Gligoroski and Panny recalled in Section~\ref{sec:introduction}, where the existence of an efficient attack on arbitrary entropic groupoids is left open.
	Our attacks sidestep this question, since they do not rely on a structure theorem and work directly in the centralizer.

	These observations suggest the following directions for future work.
	\begin{itemize}
		\item \emph{Requirements for a secure instance.} It is not enough that recovering a secret is hard: interpolation or division must be hard for every efficiently representable family of operations in the centralizer of the public data.
			Since the centralizer grows when the public operations are few or simple (Example~\ref{ex:matrices}), this points towards entropic algebras that are far from affine.
			Making this requirement precise is a first open problem.
		\item \emph{Candidate platforms.} Natural candidates are centers of multicategories, which are commutative operads by a theorem of Tronin~\cite{tronin2011natural}, and finite entropic algebras.
			Whether any such instance resists the attacks presented here remains open.
		\item \emph{Measuring hardness computationally.} On finite entropic algebras the interpolation problem is a constraint satisfaction problem.
			SAT, constraint programming and integer programming solvers offer a practical way to measure its hardness on concrete instances.
		\item \emph{Other platforms.} We have not addressed the groupoid platform of~\cite[Section 3]{gaynullina2016platforms}, which does not involve operads.
	\end{itemize}

	\section*{Acknowledgements}

	This work grew, several years later, out of a question left open in the author's master thesis~\cite{bernardelli2021operad}, which described operad-based cryptographic protocols without analyzing them.
	The author is grateful to Prof.\ Alberto Canonaco (University of Pavia), who supervised that thesis and first introduced the author to operads.

	\bibliography{references}

\end{document}